\documentclass[12pt]{iopart}
\usepackage{graphicx} % needed for figures
\usepackage[english]{babel}
\usepackage[utf8]{inputenc}
\usepackage{xcolor}
\begin{document}

\title{Using smartphones as hydrophones: two experiments in underwater acoustics}
\author{Martín Monteiro$^{1,2}$, Arturo C. Mart{\'i}$^2$ }

\address{$^1$ Universidad ORT Uruguay}
\address{$^2$ Instituto de F\'{i}sica, 
  Universidad de la Rep\'{u}blica,
 Uruguay}

\ead{marti@fisica.edu.uy}

\date{\today}
\begin{abstract}
The use of smartphone microphones in aquatic media  is explored by means 
of two experiments.  The first experiment consists in a simple time-of-flight measurement of the sound speed in water while the second deals with the acoustic location --or ranging-- of a distant object. As the underwater noise is considerable, the experimental details and the uncertainties are worth discussing.
\end{abstract}

\maketitle

\textbf{Underwater acoustics.} 
During the last years, it has become increasingly clear that smartphones are valuable tools to be used almost everywhere. Until recently, a place that still resisted smartphone onslaught was the aquatic media. 
Several experiments in acoustic were proposed 
\cite{vogt2012determining,yavuz2015measuring,kasper2015stationary,hirth2015measurement,monteiro2015measuring,monteiro2018bottle,staacks2019simple,Hellesund_2019}.
However, nowadays, many modern smartphones are waterproof and the performance of their microphones results sufficiently adequate to employ them as hydrophones. This capability gives rise to several interesting applications. Here, we describe two experiments in underwater acoustics which require two smartphones --at least one should be waterproof. The first experiment consists in a simple time-of-flight measurement of the sound speed in water and the comparison with the corresponding value in air. The second experiment deals with the acoustic location --or ranging-- of a distant object by comparing the time it takes for the sound to reach the object traveling in two different media (air and water in this case) with known sound speed. 

\begin{figure}[h]
\centering
\includegraphics[width=.5\columnwidth]{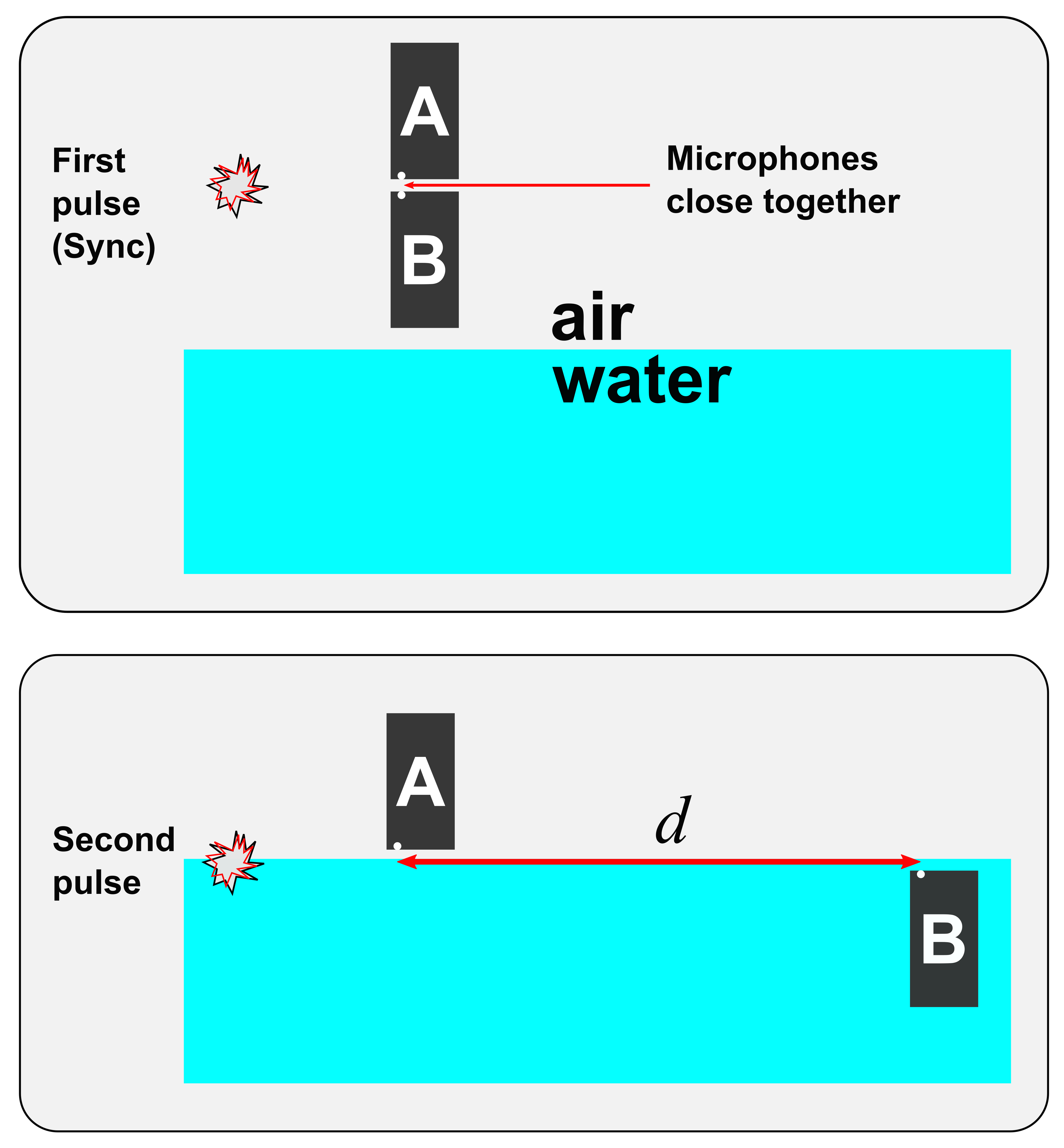}
\caption{Experiment to measure sound speed in water. Firstly, depicted the top panel, with both smartphones as close as possible, sound tracks are synchronized. Then, shown in the bottom panel, a sound pulse is registered by both smartphones separated a known distance $d$. 
}
\label{fig01}
\end{figure}

\textbf{Sound speed in water.}
A first experiment using the smartphone hydrophone, schematized in Fig.~\ref{fig01}, is to determine the sound speed in water. Although the idea is simple, care should be taken to avoid uncertainties coming from underwater noise. Two smartphones, \textit{A} and \textit{B} (at least one waterproof) and a simple app able to register and edit the raw uncompressed sound wave are needed. First, to synchronize the recordings, while both smartphones are in air recording sound with their microphones as close as possible, a sound pulse is generated (top panel). Without stopping the recording, the smartphone \textit{B} is submerged into water and separated from \textit{A} a distance $d$. Then, a second sound pulse is generated at the surface of the water. It is important that the sound source is aligned with the smartphones, so that the sound wave arrives first to smartphone \textit{A} and then to \textit{B} (bottom panel). After stopping the recordings, the audio files are saved in the cloud to further process on a computer.

\begin{figure}[h]
\centering
\includegraphics[width=.67\columnwidth]{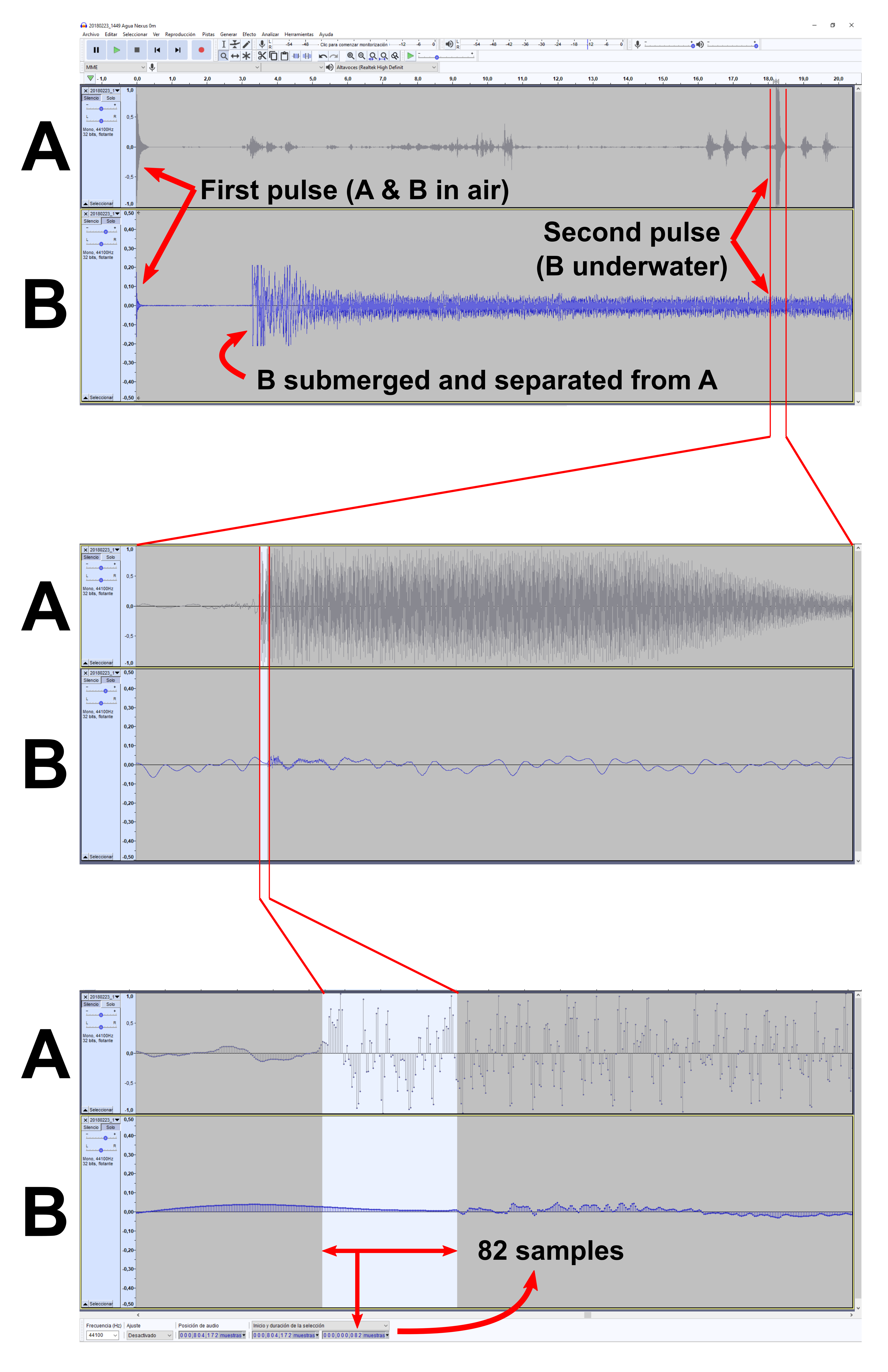}
\caption{Sounds recorded by the two smartphones displayed on the software Audacity (full track and two successive zooms). The upper (lower) track is the sound recorded by smartphone A (B). The tracks has been moved so that the first pulse matches both recordings. The second pulse occurs when the two smartphones are separated, so that it first reaches A and then B.
}
\label{fig02}
\end{figure}

The results are analyzed in Fig.~\ref{fig02}, which displays the sound waves as shown in the free software Audacity. Top and bottom audio tracks correspond to smartphones \textit{A} and \textit{B} respectively. The considerable level of underwater noise can be appreciated in the smartphone \textit{B} sound wave. The time difference in the arriving of the pulses --measured in the progressive zooms displayed in Fig.~\ref{fig02} --corresponds to 82 samples and its uncertainty is estimated in $4$ samples. As the sampling frequency is $44.1kHz$, then, the time difference results $1.9(1)ms$

The distance is directly measured underwater as $d=2.95(2) m$. Its uncertainty comes from the size of the objects (source and hydrophone) and difficulty to measure underwater. Finally, the sound speed in water will simply be the ratio between the distance, and the time it speed in water results
$$c_{water}=1.6(1) km/s$$
with very good concordance with reference values at the temperature,, at the moment of the experiment. It is also worth pointing out that sound propagates four times faster in water than in air. 

\textbf{Acoustic ranging.} A well-known problem is to determine the distance at which a sound source is located using the difference in propagation time of an acoustic pulse in two different media. For example, suppose a whale is sighted from a ship. Sound waves travel both through water and air, at different speed, and registered on board using a hydrophone and a microphone. From the time difference of the arrival time between both signals the distance to the source, x, can be calculated. The time difference between both media is easily obtained as  $ \Delta t= t_{air} - t_{water}= x 
\left( 1/c_{air} - 1/c_{water} \right)$
and the distance results
$$
  x= \frac{\Delta t}{ \frac 1{c_{air}} - \frac{1}{c_{water}} }.
$$

\begin{figure}[h]
\centering
\includegraphics[width=.6\columnwidth]{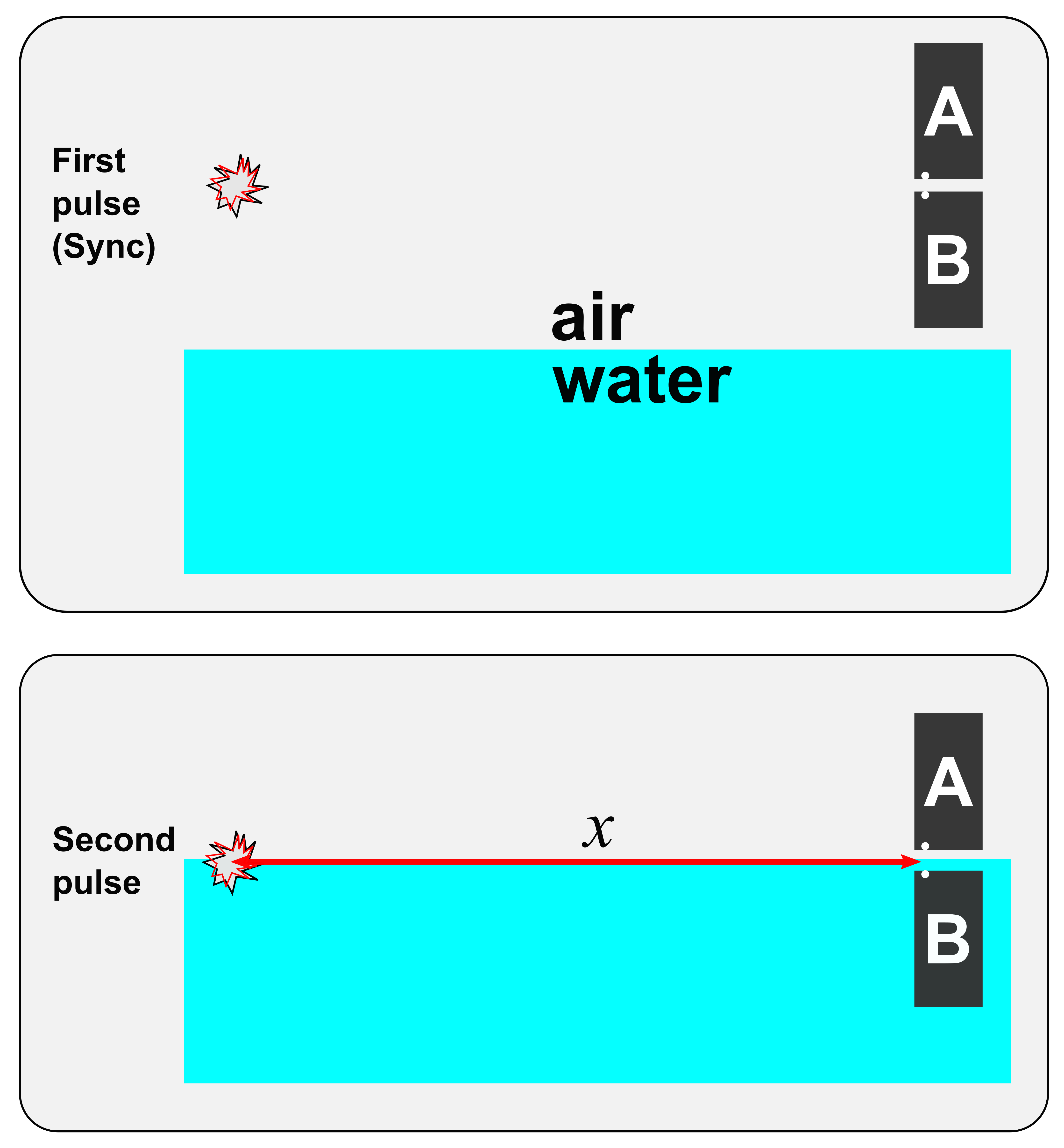}
\caption{Acoustic ranging. Similarly to the previous experiment, sound tracks are synchronized as schematized in the top panel. Then, shown in the bottom panel, a sound pulse, propagating both in water and air, reaches the smartphones separated an unknown distance $x$. 
}
\label{fig03}
\end{figure}

With a submersible smartphone (hydrophone) placed inside the water and another smartphone outside the water --very close to the first-- (microphone), the above method can be tested as shown in Fig.~\ref{fig03}. As in the previous experiment the first step is to synchronize the sound tracks. In a home pool, a pulse was produced with a semi-submerged metal bar and the time difference detected between the hydrophone and the microphone. In the present experiment, the time difference measured was 258 samples, equivalent to  $5.9(1)  ms$. Then, the distance from the metal bar to the smartphones is readily obtained as $x=2.66(6) m$. This result is in very good concordance with the direct measurement of the distance, $x_d=2.60(2) m$ between the smartphones .

\textbf{Closing remarks.} To sum up, we have described a simple and inexpensive experiment in underwater acoustics, an area usually little explored in Physics courses. At least one waterproof smartphone is needed and the experiments can be performed in a small pool or pond. Data analysis requires only a sound editor, however, due to the noise inherent to the aquatic media, special attention should be paid to the experimental details. 

\textbf{Acknowledgements} We acknowledge financial support from CSIC (UdelaR, Uruguay) and Programa de Desarrollo de las Ciencias Basicas (Uruguay).

\section*{References}

\bibliography{/home/arturo/Dropbox/bibtex/mybib}

\begin{thebibliography}{1}

\bibitem{vogt2012determining}
Patrik Vogt and Jochen Kuhn.
\newblock Determining the speed of sound with stereo headphones.
\newblock {\em The Physics Teacher}, 50(5):308--309, 2012.

\bibitem{yavuz2015measuring}
Ahmet Yavuz.
\newblock Measuring the speed of sound in air using smartphone applications.
\newblock {\em Physics Education}, 50(3):281, 2015.

\bibitem{kasper2015stationary}
Lutz Kasper, Patrik Vogt, and Christine Strohmeyer.
\newblock Stationary waves in tubes and the speed of sound.
\newblock {\em The Physics Teacher}, 53(1):52--53, 2015.

\bibitem{hirth2015measurement}
Michael Hirth, Jochen Kuhn, and Andreas M{\"u}ller.
\newblock Measurement of sound velocity made easy using harmonic resonant
  frequencies with everyday mobile technology.
\newblock {\em The Physics Teacher}, 53(2):120--121, 2015.

\bibitem{monteiro2015measuring}
Mart{\'\i}n Monteiro, Arturo~C Marti, Patrik Vogt, Lutz Kasper, and Dominik
  Quarthal.
\newblock Measuring the acoustic response of helmholtz resonators.
\newblock {\em The Physics Teacher}, 53(4):247--249, 2015.

\bibitem{monteiro2018bottle}
Martín Monteiro, Cecilia Stari, Cecilia Cabeza, and Arturo~C. Marti.
\newblock A bottle of tea as a universal helmholtz resonator.
\newblock {\em The Physics Teacher}, 56(9):644--645, 2018.

\bibitem{staacks2019simple}
Sebastian Staacks, Simon H{\"u}tz, Heidrun Heinke, and Christoph Stampfer.
\newblock Simple time-of-flight measurement of the speed of sound using
  smartphones.
\newblock {\em The Physics Teacher}, 57(2):112--113, 2019.

\bibitem{Hellesund_2019}
Simen Hellesund.
\newblock Measuring the speed of sound in air using a smartphone and a
  cardboard tube.
\newblock {\em Physics Education}, 54(3):035015, apr 2019.

\end{thebibliography}

\bibliographystyle{unsrt}

\end{document}